\documentclass[showpacs,epsf,twocolumn]{revtex4-1}
\usepackage{float}
\usepackage{color,graphicx}
\graphicspath{ {images/} }
\usepackage{hyperref}
\begin{document}

\title{Planar Hall effect in type II Dirac semimetal VAl$_{3}$}

\author{Ratnadwip Singha, Shubhankar Roy, Arnab Pariari, Biswarup Satpati, Prabhat Mandal}

\email{prabhat.mandal@saha.ac.in}

\affiliation{Saha Institute of Nuclear Physics, HBNI, 1/AF Bidhannagar, Calcutta 700 064, India}

\date{\today}

\begin{abstract}

The study of electronic properties in topological systems is one of the most fascinating topics in condensed matter physics, which has generated enormous interests in recent times. New materials are frequently being proposed and investigated to identify their non-trivial band structure. While sophisticated techniques such as angle-resolved photoemission spectroscopy have become popular to map the energy-momentum relation, the transport experiments lack any direct confirmation of Dirac and Weyl fermions in a system. From band structure calculations, VAl$_{3}$ has been proposed to be a type II topological Dirac semimetal. This material represents a large family of isostructural compounds, all having similar electronic band structure and is an ideal system to explore the rich physics of Lorentz symmetry violating Dirac fermions. In this work, we present a detailed analysis on the magnetotransport properties of VAl$_{3}$. A large, non-saturating magnetoresistance has been observed. Hall resistivity reveals the presence of two types of charge carriers with high mobility. Our measurements show a large planar Hall effect in this material, which is robust and can be easily detectable up to high temperature. This phenomenon originates from the relativistic chiral anomaly and non-trivial Berry curvature, which validates the theoretical prediction of the Dirac semimetal phase in VAl$_{3}$.
\end{abstract}

\maketitle

The quest for novel quantum states of matter has yielded the recent discovery of non-trivial topological phases and reshaped our understanding of condensed matter physics. The unconventional and vastly unexplored physics of these materials has triggered a whole new area of research, which has flourished in last few years. Topological insulators \cite{Zhang,Xia,Chen}, Dirac/Weyl semimetals \cite{Liu1,Liu2,Lv,Xu1} and topological superconductors \cite{Xu2,Matano} have enabled us to explore the dynamics of relativistic particles such as Dirac, Weyl and Majorana fermions in low energy electronic systems. On the other hand, from unique combinations of different crystalline symmetries, new particles have emerged, which are impossible to realize in high energy physics \cite{Soluyanov,Yan}. Type I Dirac (Weyl) fermions appear as quasiparticle excitations at the crossing point of the fourfold (twofold) degenerate linear electronic bands in bulk state of three dimensional topological semimetals (TSMs) \cite{Liu1,Liu2,Lv,Xu1}. In type II Dirac/Weyl semimetals, however, an additional kinetic term in the energy spectrum tilts the Dirac/Weyl cone and as a result the Lorentz invariance breaks \cite{Soluyanov,Yan}. While preserving the Lorentz symmetry is one of the fundamental conditions in quantum field theory, condensed matter systems do not have this constraint. Thus, these topological materials provide the sole opportunity to study Lorentz-violating Dirac and Weyl fermions, which are predicted to exhibit several unconventional physical phenomena \cite{Udagawa,Yu,OBrien}.

Although type II Weyl fermions have been probed in few systems \cite{Soluyanov,Deng,Koepernik,Chang,Belopolski}, the candidates for type II Dirac semimetals are rather limited \cite{Yan,Fei}. Recently, from band structure calculations, the members of the family \textit{MA$_{3}$} ($M$ = V, Nb, Ta; $A$ = Al, Ga, In) are proposed to be type II Dirac semimetals \cite{Chang2}. The Dirac cone appears at the contact point of electron and hole pockets and shows a mirror Chern number 2, which so far, has not been observed in type I TSMs. Under broken time-reversal symmetry, this Dirac node splits into two quadratic Weyl nodes carrying chiral charges $\pm$2. Each quadratic Weyl cone can be further broken into two linearly dispersing Weyl cones, once the inherent $C_{4}$ rotational symmetry of the crystal structure is lifted. Similarly, the material can be driven to a topological crystalline insulating state by only breaking the $C_{4}$ rotational symmetry. Hence, \textit{MA$_{3}$} family offers an excellent prospect and tunability to investigate type II Dirac fermions as well as other exotic topological phases.

Identifying Dirac or Weyl fermions in a material is a formidable task and requires sophisticated techniques like angle-resolved photoemission spectroscopy (ARPES). Nevertheless, the signatures of Dirac/Weyl fermions can be observed in transport experiments. Large magnetoresistance (MR), ultra high carrier mobility and small carrier effective mass are some of the characteristics of the TSMs and considered as general criteria to predict new compounds in this class \cite{Ali,Liang,Shekhar,Singha1}. However, the presence of parabolic bands near Fermi energy often hinder these prominent features. On the other hand, different mechanisms such as electron-hole compensation \cite{Yuan} or the presence of open-orbit Fermi surface \cite{Singha2} can also show the above-mentioned properties. In addition, multiple Fermi pockets make it difficult to accurately extract the Berry phase from quantum oscillation. The Adler-Bell-Jackiw (ABJ) chiral anomaly is widely regarded as a direct evidence of the Dirac/Weyl cones in the band structure \cite{Huang,Li,Li2,Wang}. This relativistic effect is generated from the non-conservation of the number of Weyl fermions with a definite chirality and results in negative MR under parallel electric and magnetic fields \cite{Huang,Li}. To observe this phenomenon, the primary condition is to have the Dirac or Weyl nodes very close to the Fermi energy \cite{Li2,Wang}, which is not the case for most TSMs. As these materials show large positive MR at transverse electric and magnetic field configuration, a small misalignment between them can also easily mask the weak negative MR component \cite{Huang}. Moreover, other mechanisms such as current jetting \cite{Hu}, weak localization \cite{Ulmet} can also produce negative MR. Therefore, it is very difficult to distinguish the negative MR induced by chiral anomaly from others. From theoretical calculations, recently, a planar Hall effect (PHE) has been predicted in TSMs, which originates from the ABJ chiral anomaly and non-trivial Berry curvature \cite{Burkov,Nandy}. As PHE is completely different from usual Hall effect, both in experimental configuration and angle dependence, this phenomenon is easier to identify and unambiguously confirms the Dirac/Weyl type excitations in the system. Large PHE has already been reported in topological insulator ZrTe$_{5}$ \cite{Li3}, Dirac semimetal Cd$_{3}$As$_{2}$ \cite{Li4} and Weyl semimetals WTe$_{2}$ \cite{Wang2}, GdPtBi \cite{Kumar}. It is worth mentioning that a weak PHE can be observed in ferromagnetic compounds due to magnetic ordering \cite{Nazmul}. However, as most of the TSMs are non-magnetic, this effect does not arise in these systems.

In this paper, we report the magnetotransport properties of newly proposed type II Dirac semimetal VAl$_{3}$.  Large, non-saturating MR has been observed along with large carrier mobility. The non-linear field dependence of Hall resistivity confirms the presence of both electron and hole type carriers, as predicted from theoretical calculations. A prominent PHE has been observed, which persists up to high temperature and reveals the non-trivial nature of the band structure in this material.

The single crystals of VAl$_{3}$ were grown in aluminium-flux \cite{Canfield}. High purity V granules (Alfa Aesar 99.8\%) and Al pieces (Alfa Aesar 99.999\%) were taken in an alumina crucible in 1:9 molar ratio. The crucible was put in a quartz tube, which was then evacuated and sealed. The quartz tube was heated to 950$^\circ$C and soaked at this temperature for 12 hrs. After that the tube was slowly cooled (3$^\circ$C/h) to 750$^\circ$C. At this temperature, the excess aluminium was decanted using a centrifuge. Several needle like single crystals with typical dimensions 1.5$\times$0.3$\times$0.2 mm$^{3}$ were obtained, which were further etched in sodium hydroxide solution to remove any remaining aluminium on the surface. The phase purity was confirmed by x-ray diffraction (XRD) of the powdered single crystals in Rigaku TTRAX III diffractometer using Cu K$\alpha$ radiation. The high resolution transmission electron microscopy (HRTEM) was done in an FEI, TECNAI G$^{2}$ F30, S-TWIN microscope operating at 300 kV and equipped with a GATAN Orius SC1000B CCD camera. The elemental composition was checked by energy dispersive x-ray (EDX) spectroscopy in the same microscope. The transport measurements were performed by four probe technique using a Quantum Design 9 T physical property measurement system (PPMS) with horizontal sample rotator option. The electrical contacts were made with gold wire and conducting silver paint.

VAl$_{3}$ crystallizes in a tetragonal structure with crystallographic symmetry \textit{I}4/\textit{mmm} (space group no. 139) \cite{Chang2,Maas}. As evident from the crystal structure [Fig. 1(a) inset], each Al atom is surrounded by four neighboring V atoms either in planar square or tetrahedron geometry. The image of a typical single crystal is shown in Fig. S1 with appropriate length scales \cite{Supp}. Fig. 1(a) represents the XRD pattern of the powdered crystals. The experimental data have been analyzed by Rietveld structural refinement using FULLPROF software package. The refined lattice parameters, $a$=$b$=3.777(2) and $c$=8.324(5) {\AA}, are in excellent agreement with earlier report \cite{Maas}. Within our experimental resolution, the absence of any impurity peak, confirms the phase purity of the grown single crystals. The HRTEM image of a single crystal along \textit{ab} plane is shown in Fig 1(b), which reveals the high quality crystalline nature of the sample. The calculated interplanar spacing 3.52(4) {\AA} is consistent with the extracted lattice parameters from XRD data. To check the elemental composition, we have performed EDX spectroscopy on the as grown single crystals. From the experimental data [Fig. 1(c)], the V and Al atomic ratio is found to be 1:2.8 with maximum relative error $\sim$5\%, which is close to the desired stoichiometry. The C and Cu signals in EDX spectrum correspond to the carbon coated copper grid on which the sample was mounted for HRTEM.

\begin{figure}
\includegraphics[width=0.5\textwidth]{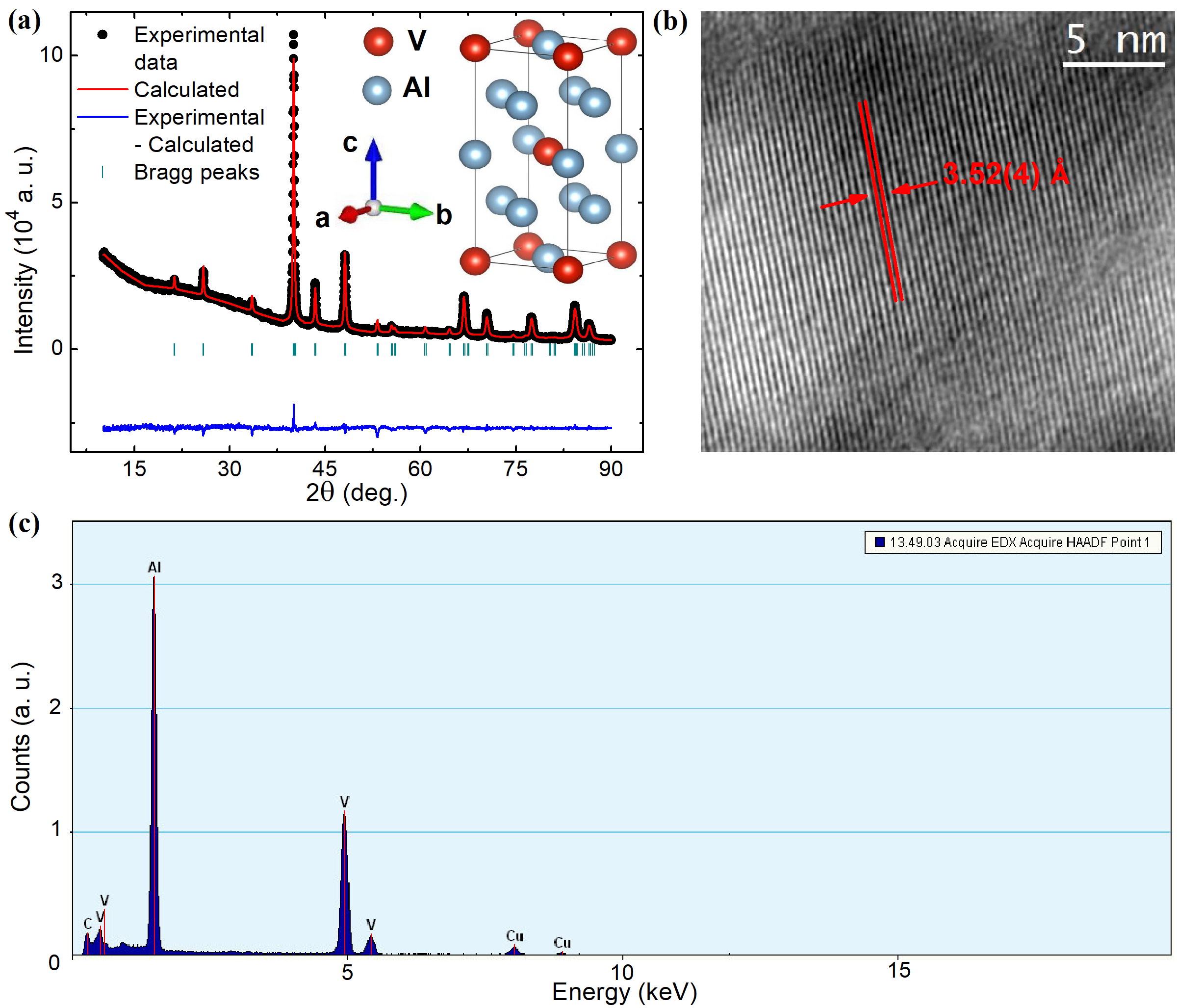}
\caption{(a) X-ray diffraction spectra of the powdered single crystals. The experimental data have been analyzed by Rietveld structural refinement. Inset shows the crystal structure of VAl$_{3}$. (b) The high-resolution transmission electron microscopy image of the as grown crystal. (c) Energy dispersive x-ray spectroscopy results of a typical single crystal.}
\end{figure}

In Fig. 2(a), the longitudinal resistivity ($\rho_{xx}$) of a typical single crystal is plotted as a function of temperature. VAl$_{3}$ exhibits metallic character, i.e., $\rho_{xx}$ decreases monotonically from room temperature. At 2 K, the resistivity becomes $\sim$9.3 $\mu\Omega$ cm and gives a residual resistivity ratio [$\rho_{xx}$(300 K)/$\rho_{xx}$(2 K)] $\sim$12. As shown in the inset, in the low temperature region, $\rho_{xx}$ follows a quadratic temperature dependence, which is consistent with the Fermi liquid theory \cite{Ziman}. Above 100 K, $\rho_{xx}$ increases linearly with temperature. Under an applied magnetic field perpendicular to the current, the low-temperature resistivity increases. However, within the applied magneic field range, we have not observed any field-induced metal-semiconductor like crossover as in most TSMs \cite{Singha1,Shekhar,Huang,Ali}. Next, we have measured the magnetic field dependence of the resistivity at different temperatures [Fig. 2(b)]. At transverse magnetic field and current configuration, a large, non-saturating MR [$\frac{\rho(B)-\rho(0)}{\rho(0)}$$\times$100 \%] $\sim$158 \% has been observed at 2 K and 9 T. This value is two to three orders of magnitude smaller than that for a typical TSM \cite{Liang,Singha1,Shekhar,Huang,Ali}. The band structure of VAl$_{3}$ consists of several parabolic bands, crossing the Fermi energy \cite{Chang2}. Hence, it is expected that the transport properties are dominated by the carriers from these trivial Fermi pockets. With increasing temperature, MR decreases drastically and becomes negligible at 300 K. In the inset of Fig. 2(b), we have fitted $\rho_{xx}(B)$ by a $\alpha B^{m}$ type power-law with $m\sim$1.3, where $\alpha$ is an arbitrary parameter. From semiclassical two-band theory, a quadratic field dependence is predicted for compensated semimetals with equal density of electron and hole type carriers \cite{Ziman}. For VAl$_{3}$, however, $m\neq$2 indicates an uncompensated scenario.

\begin{figure}
\includegraphics[width=0.5\textwidth]{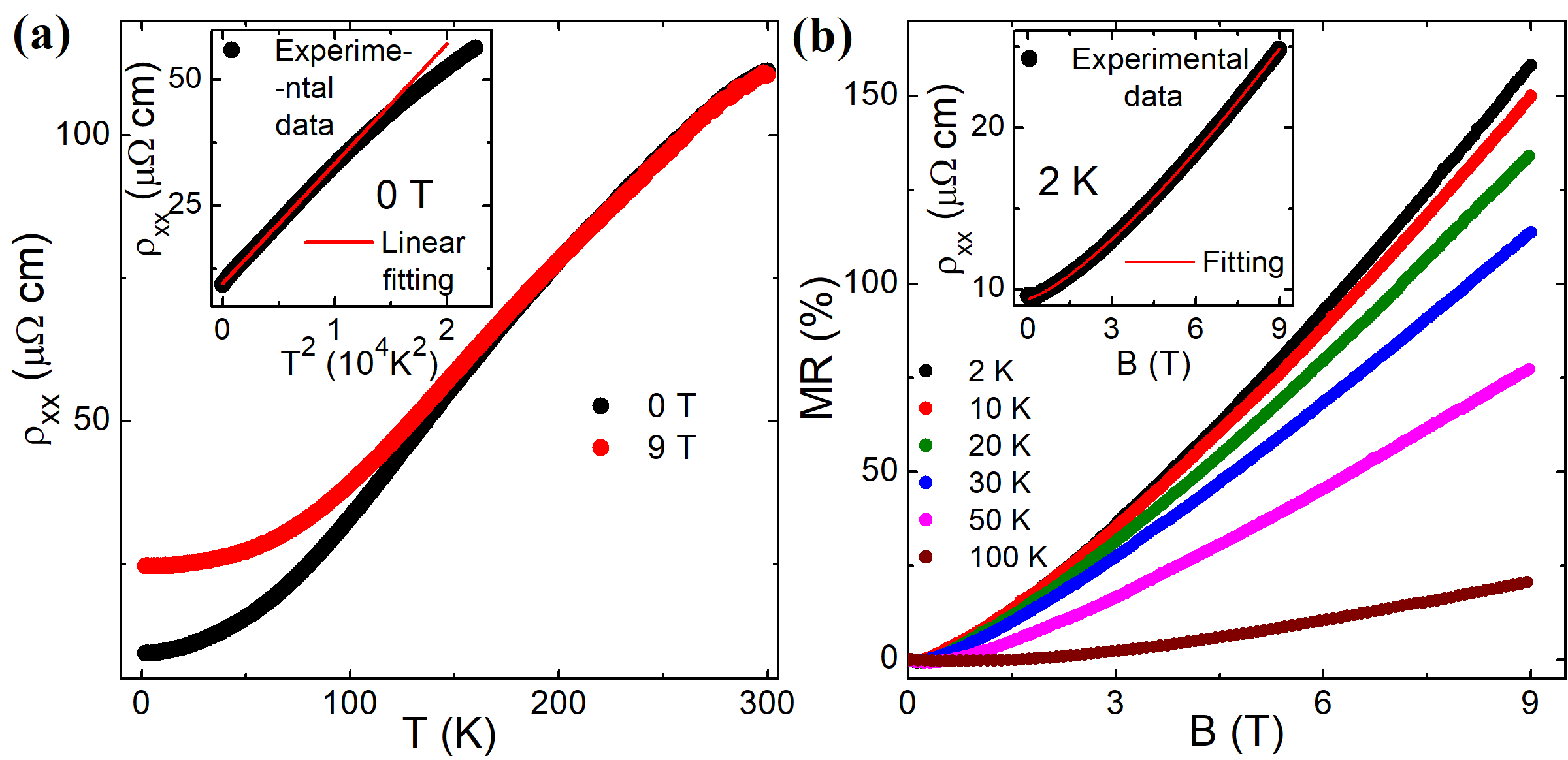}
\caption{(a) Temperature dependence of the longitudinal resistivity ($\rho_{xx}$) without and with a transverse magnetic field. The inset manifests the $T^{2}$ dependent behavior of $\rho_{xx}$ at low temperature. (b) Transverse magnetoresistance of VAl$_{3}$ at different temperatures. In the inset, the experimental data have been fitted using $\alpha B^{m}$-type power-law.}
\end{figure}

To obtain further information about the carrier density and mobility of the carriers, we have performed the Hall resistivity measurement. In Fig. 3(a) the Hall resistivity ($\rho_{yx}$) is plotted as a function of magnetic field at few representative temperatures. At 300 K, $\rho_{yx}(B)$ is almost linear and negative, implying that the majority carrier is electron type. With decreasing temperature, a non-linear behavior appears and persists down to 2 K. The non-linear field dependence of Hall resistivity confirms the presence of more than one type of carrier and different temperature dependence of their mobilities. Using the tensorial inversion of the resistivity matrix, we have calculated the Hall conductivity $\sigma_{xy}$=$\frac{\rho_{yx}}{\rho_{yx}^{2}+\rho_{xx}^{2}}$ and shown in Fig. 3(b). The experimental data have been fitted [Fig. 3(b) inset] using semiclassical two-band theory \cite{Hurd},
\begin{equation}
\sigma_{xy}=[n_{h}\mu_{h}^{2}\frac{1}{1+(\mu_{h}B)^2}-n_{e}\mu_{e}^{2}\frac{1}{1+(\mu_{e}B)^2}]eB,
\end{equation}
where $n_{h}$ ($n_{e}$) and $\mu_{h}$ ($\mu_{e}$) are the hole (electron) density and mobility, respectively. From the fitting parameters, the extracted electron and hole densities are 7.7(3)$\times$10$^{19}$ and 3.7(1)$\times$10$^{19}$ cm$^{-3}$, respectively at 2 K. Similarly, at 2 K, the mobility of the electrons and holes are found to be 4.8(1)$\times$10$^{3}$ and 0.3(2)$\times$10$^{3}$ cm$^{2}$V$^{-1}$s$^{-1}$, respectively. These values manifest that the magnetotransport properties in VAl$_{3}$ are primarily influenced by electron-type charge carrier. The obtained carrier density and mobility are comparable to those for several TSMs \cite{Novak,Pavlosiuk,Luo}.

\begin{figure}
\includegraphics[width=0.5\textwidth]{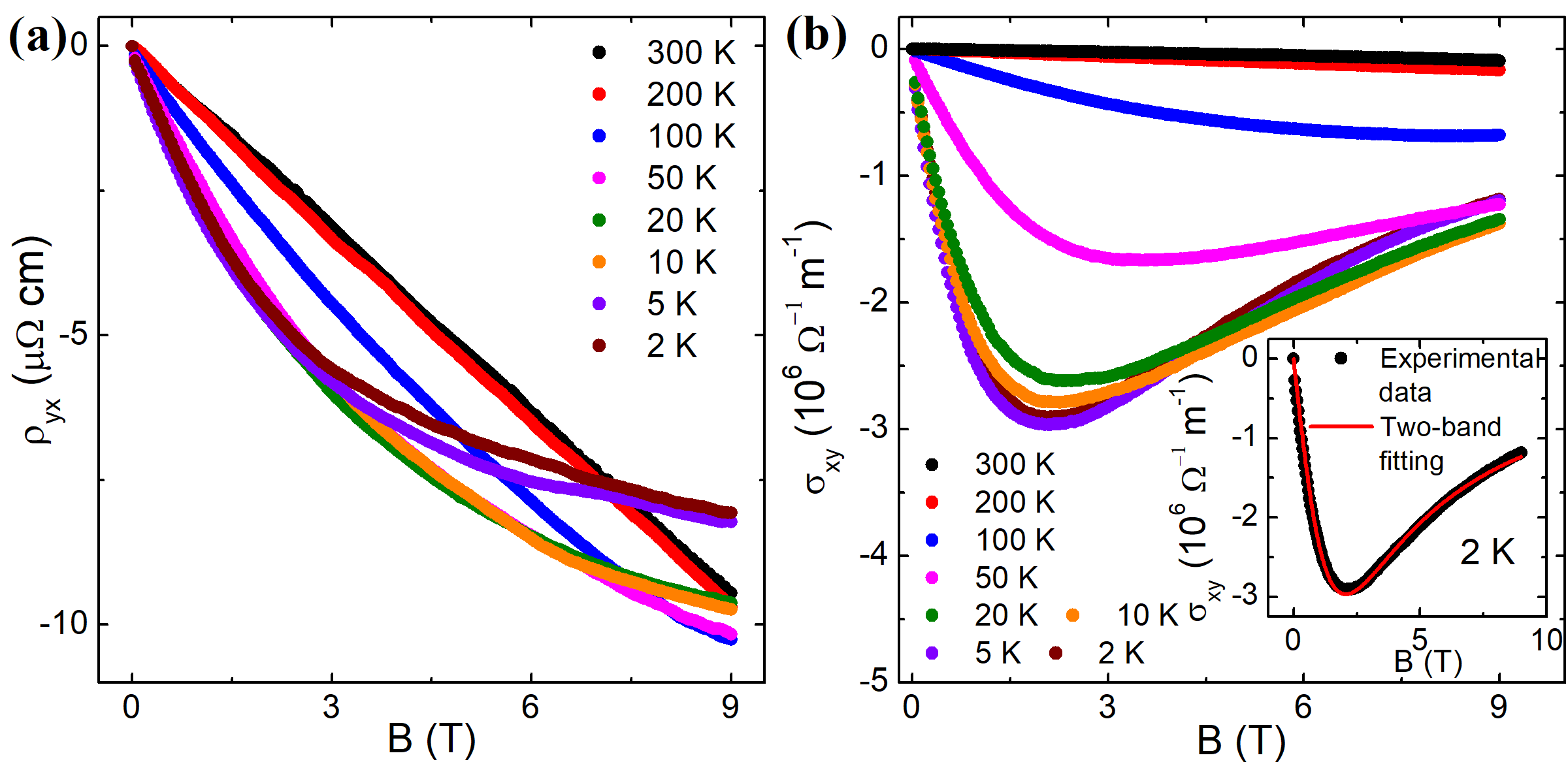}
\caption{(a) Hall resistivity ($\rho_{yx}$) as a function of magnetic field at different temperatures. (b) Field dependence of the calculated Hall conductivity ($\sigma_{xy}$=$\frac{\rho_{yx}}{\rho_{yx}^{2}+\rho_{xx}^{2}}$). Inset shows the two-band fitting of $\sigma_{xy}$ at 2 K as a representative.}
\end{figure}

\begin{figure}
\includegraphics[width=0.5\textwidth]{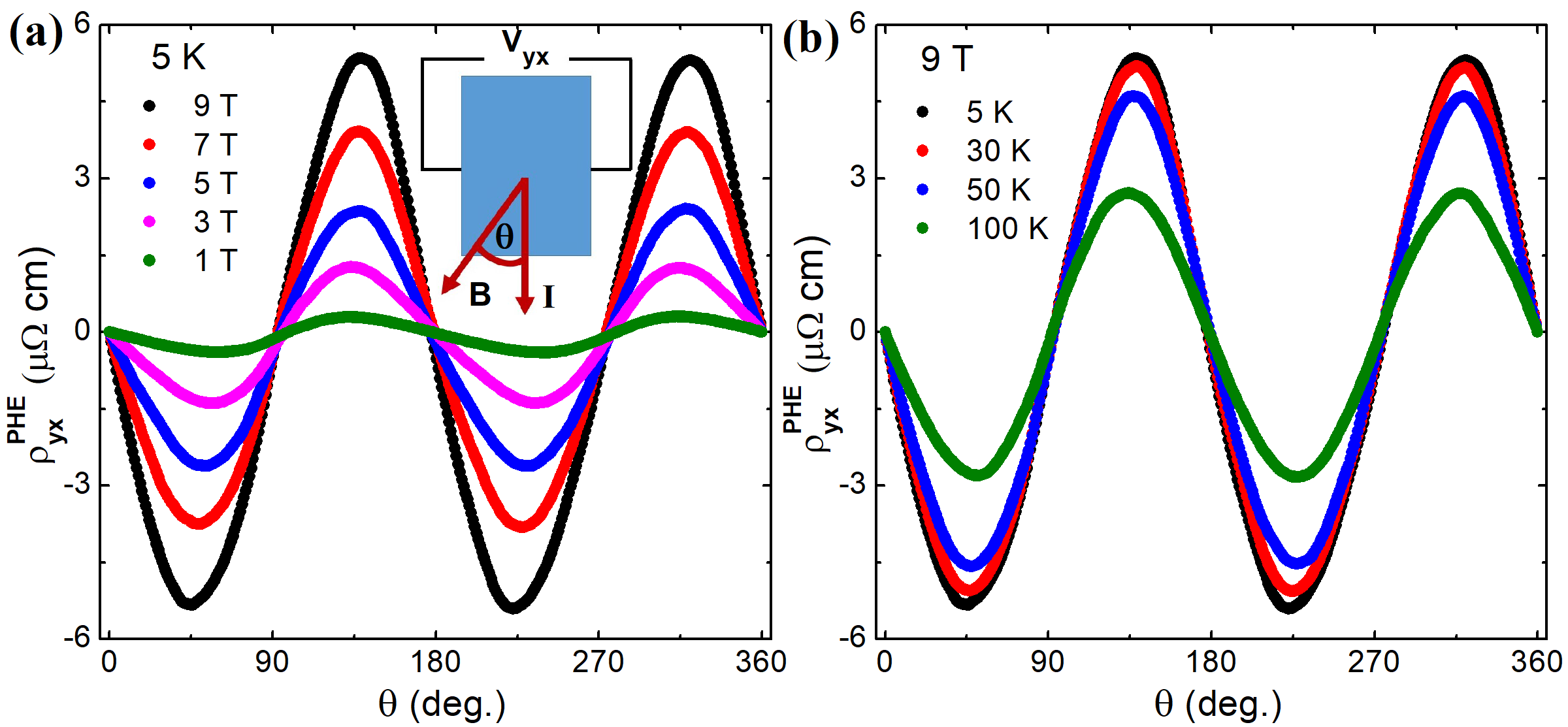}
\caption{(a) Angle dependence of the planar Hall resistivity ($\rho_{yx}^{PHE}$) at 5 K for different magnetic field strength. Inset shows the schematic for experimental set-up. (b) Angle dependence of $\rho_{yx}^{PHE}$ at 9 T for different temperatures.}
\end{figure}

Identifying the topological nature of the band structure in VAl$_{3}$ is of fundamental interest to confirm the theoretical prediction of the type II Dirac fermions in this material \cite{Chang2}. However, the complex band structure with several bands coexisting at the Fermi energy \cite{Chang2}, makes it very difficult to probe the Dirac cone from low-energy experiments. Due to this reason, recent quantum oscillation study on the members of the \textit{MA$_{3}$} family could not confirm the Berry phase for different Fermi pockets \cite{Chen2}. PHE, on the other hand, does not possess such shortcoming and is a direct evidence of the Dirac/Weyl fermions in the system \cite{Burkov,Nandy}. In the inset of Fig. 4(a), we have shown the experimental set-up for PHE measurement, where the current, magnetic field and the Hall voltage ($V_{yx}$) are coplanar. During the measurement, magnetic field is rotated within this plane; $\theta$ being the angle between the current and magnetic field. In Fig. 4(a), we have shown the angle dependence of the planar Hall resistivity ($\rho_{yx}^{PHE}$) for VAl$_{3}$ at 5 K for different magnetic field strengths. To exclude any contribution from the conventional Hall voltage, we did the measurements with both positive and negative field directions and took the average. $\rho_{yx}^{PHE}$ shows a periodic behavior with the amplitude decreasing with decreasing magnetic field. The most striking aspect of PHE is the position of the extrema points. Due to the very nature of the Lorenz force, the Hall resistivity is zero for an applied magnetic field parallel to the current, whereas it becomes extremum when magnetic field, current and Hall voltage are mutually perpendicular for conventional Hall measurements. In Fig. S2, we have plotted the conventional Hall resistivity of VAl$_{3}$ as a function of the angle between the current and magnetic field \cite{Supp}. As expected, $\rho_{yx}$ shows maxima and minima at 90$^{\circ}$ and 270$^{\circ}$, respectively. In PHE, the experimental configuration ensures that there is no Lorenz force. As evident from Fig. 4(a), the experimental data show minima and maxima at 45$^{\circ}$ (and 225$^{\circ}$) and 135$^{\circ}$ (and 315$^{\circ}$), respectively. These positions of extrema are consistent with the theoretically predicted PHE \cite{Nandy} and confirm the non-trivial nature of the band structure in VAl$_{3}$. PHE in topological systems can be mathematically formulated from the semiclassical Boltzmann theory \cite{Burkov,Nandy},
\begin{equation}
\rho_{yx}^{PHE}=-\Delta\rho_{chiral} \sin\theta \cos\theta
\end {equation}
\begin{equation}
\rho_{xx}^{planar}=\rho_{\perp}-\Delta\rho_{chiral}\cos^{2}\theta
\end {equation}

\begin{figure}
\includegraphics[width=0.5\textwidth]{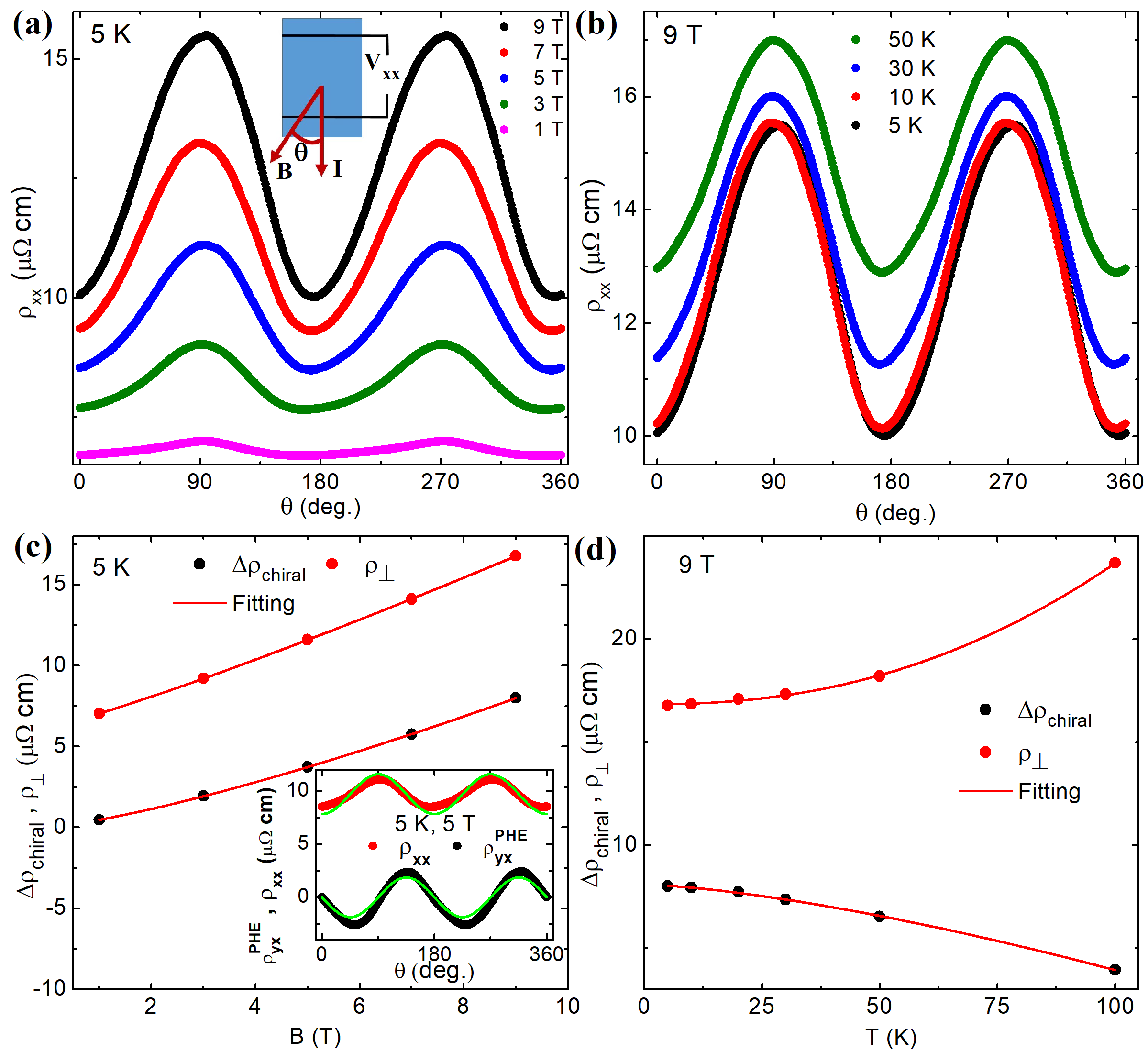}
\caption{(a) Anisotropy in planar resistivity ($\rho_{xx}^{planar}$) for different magnetic field. The measurement set-up is shown schematically in the inset. (b) Angle dependence of $\rho_{xx}^{planar}$ at different temperatures. (c) The magnetic field dependence of the extracted chiral anomaly induced resistivity component and transverse resistivity. Inset shows the global fitting of $\rho_{yx}^{PHE}$ and $\rho_{xx}^{planar}$ using Eqs. (2) and (3). (d) The temperature dependence of $\Delta\rho_{chiral}$ and $\rho_{\perp}$ at 9 T.}
\end{figure}

where $\Delta\rho_{chiral}=\rho_{\perp}-\rho_{\parallel}$ is the chiral anomaly induced resistivity component and $\rho_{xx}^{planar}$ corresponds to the planar resistivity. $\rho_{\perp}$ and $\rho_{\parallel}$ are the resistivity for current perpendicular (transverse resistivity) and parallel to the magnetic field direction, respectively. In Fig. 4(b) we have plotted $\rho_{yx}^{PHE}$ for different temperatures at applied magnetic field 9 T. $\rho_{yx}^{PHE}$ shows weak temperature dependence and remains quite large and detectable at 100 K and above. This is in sharp contrast to the chiral anomaly induced negative MR, which vanishes quickly with increasing temperature \cite{Huang,Li,Li2,Wang}. We have also measured the planar resistivity $\rho_{xx}^{planar}$ in VAl$_{3}$ for different magnetic field and temperature combinations. The experimental set-up has been shown schematically in the inset of Fig. 5(a). In Fig. 5(a) and (b), $\rho_{xx}^{planar}$ has been plotted for different magnetic field strengths at 5 K and for different temperatures at 9 T, respectively. $\rho_{xx}^{planar}$ shows a periodicity $\pi$ with maximum at 90$^{\circ}$ and 270$^{\circ}$, i.e., when magnetic field and current are in transverse configuration. To further confirm the nature of the PHE, we have fitted $\rho_{yx}^{PHE}$ and $\rho_{xx}^{planar}$ using the theoretical Eqs. (2) and (3). As shown in the inset of Fig. 5(c), our global fitting is in excellent agreement with the experimental results. From the fitting parameters, we have extracted the chiral resistivity and transverse resistivity components. Both of these quantities increase monotonically with field and follow a $B^{n}$-type relation with $n\sim$1.3 [Fig. 5(c)]. Fig. 5(d) illustrates the temperature dependence of $\Delta\rho_{chiral}$ and $\rho_{\perp}$. As expected, at 9 T, $\Delta\rho_{chiral}$ decreases with increasing temperature. The experimental data can be described well by $\Delta\rho_{chiral}=C_{1}-C_{2}T^{1.4}$ type power law, where $C_{1}$ and $C_{2}$ are arbitrary constants. On the other hand, $\rho_{\perp}$ is found to increase almost quadratically with temperature.

In conclusion, we have analyzed the magnetotransport properties of single crystalline VAl$_{3}$. A large, non-saturating magnetoresistance has been observed along with high carrier mobility in this compound. The field dependence of Hall resistivity shows the presence of both electron and hole type charge carriers. From our measurements, a large planar Hall effect has been identified. This effect is robust and persists up to quite high temperature. As planar Hall effect appears due to the relativistic ABJ chiral anomaly and non-trivial Berry phase, it is an excellent tool to detect the presence of Dirac/Weyl fermions in a system. Hence, our study unambiguously confirms the theoretically predicted type II Dirac semimetal phase in VAl$_{3}$. This material represents a large family (\textit{MA$_{3}$}) of isostructural compounds, which are all proposed to be type II Dirac semimetals. Materials with identical crystal and electronic band structures provide a wide range of tunability of the electronic state, simply by changing the chemical composition or doping. For example, spin-orbit coupling (SOC) is an important parameter, which significantly influences the topological phases in a system. With increasing atomic number (Z) from V (Z=23) to Ta (Z=73), one can effectively change the SOC and hence can tune the electronic properties of the system. On the other hand, through doping, the chemical potential can be moved closer to the type II Dirac node and made it easily accessible by low energy experiments. Therefore, our work not only presents a detailed study on VAl$_{3}$, but also encourages further investigation on the other members of the \textit{MA$_{3}$} family.

\end{document}